\documentclass[10pt,twocolumn,aps,prl,nobibnotes,showpacs,superscriptaddress,floatfix]{revtex4-1}

\usepackage{graphicx}
\usepackage{amssymb}
\usepackage{amsmath}
\usepackage{color}
\usepackage{psfrag}
\usepackage{epsfig}
\usepackage{bbm}
\usepackage{bm}
\usepackage{hyperref}
\usepackage[normalem]{ulem}
\hypersetup{breaklinks}
\usepackage{amsthm}

\newcommand{\ket}[1]{| #1 \rangle}
\newcommand{\bra}[1]{\langle #1 |}

\newcommand{\beq}{\begin{eqnarray}}
\newcommand{\eeq}{\end{eqnarray}}

\DeclareMathOperator{\tr}{tr}

\theoremstyle{definition}
\newtheorem{theorem}{Theorem}
\newtheorem*{remark*}{Remark}

\definecolor{mygreen}{rgb}{0,0.5,0} 
\definecolor{myblue}{rgb}{0,0,0.75} 
\definecolor{mymagenta}{cmyk}{0,1,0,0.12}

\begin{document}
\title{One-to-one mapping between steering and joint measurability problems}
\author{Roope Uola}
\affiliation{Naturwissenschaftlich-Technische Fakult\"at, Universit\"at 
 Siegen, Walter-Flex-Str.\ 3, D-57068 Siegen, Germany}
\author{Costantino Budroni}
\affiliation{Naturwissenschaftlich-Technische Fakult\"at, Universit\"at 
 Siegen, Walter-Flex-Str.\ 3, D-57068 Siegen, Germany}
\author{Otfried G\"uhne}
\affiliation{Naturwissenschaftlich-Technische Fakult\"at, Universit\"at 
 Siegen, Walter-Flex-Str.\ 3, D-57068 Siegen, Germany}
\author{Juha-Pekka Pellonp\"{a}\"{a}} 
\affiliation{Turku Centre for Quantum Physics, Department of Physics and Astronomy, University of Turku, FI-20014 Turku, Finland }
\date{ \today}

\begin{abstract}
Quantum steering refers to the possibility for Alice to remotely 
steer Bob's state by performing local measurements on her half 
of a bipartite system. Two necessary ingredients for steering 
are entanglement and incompatibility of Alice's measurements. 
In particular, it has been recently proven that for the 
case of pure states of maximal Schmidt rank the problem of 
steerability for Bob's assemblage is equivalent to the problem 
of joint measurability for Alice's observables. We show that such an
equivalence holds in general, namely, the steerability of any 
assemblage can always be formulated as a joint measurability 
problem, and vice versa. We use this connection to introduce 
steering inequalities from joint measurability criteria and 
develop quantifiers for the incompatibility of measurements.
\end{abstract}

\maketitle

{\it Introduction.---}
Steering is a quantum effect by which one experimenter, Alice, can remotely prepare an ensemble of states for another experimenter, Bob, by performing local measurement on her half of a bipartite system and communicating the results to Bob. Introduced by Schr\"odinger in 1935 \cite{Schr35}, quantum steering is a form of quantum correlation intermediate between Bell nonlocality and entanglement. It has recently attracted increasing interest \cite{wiseman2007, SNC14, pusey2014,jevtic2014,Milne14,Kogias15}, both from a theoretical and experimental perspective, and it has been recognized as a resource for different tasks such as one-sided device-independent quantum key distribution \cite{Branc12,He13} and subchannel discrimination \cite{Piani15}. In addition, the question which 
quantum states can be used for steering can be addressed with efficient 
numerical techniques, contrary to the notion of entanglement or the question 
which states violate a Bell inequality. In this way, the notion of steering
has been used to find a counterexample to the Peres conjecture, 
a long-standing open problem in entanglement theory \cite{moroderperes2014, vertesiperes2014}.

A successful implementation of a steering protocol involves different elements, e.g., entangled states and incompatible measurements, and therefore steering has been investigated under different perspectives. On the one hand, allowing for an optimization over all possible quantum states or, equivalently, considering the maximal entangled state, steering has been identified with the lack of joint measurability of Alice's local observables \cite{Uola14, Quint14}, similarly to the case of nonlocality \cite{Wolf09}. On the other hand, if an optimization over all possible measurements for Alice has been considered, steering has been identified with a property of the state allowing for optimal subchannel discrimination when one is restricted to local measurements and one-way classical communication \cite{Piani15}.
In addition, a very natural and interesting framework for steering is that of one-sided device-independent (1SDI) quantum information processing. In the case of device-independent quantum information processing, both parties are untrusted, hence no assumption is made on the system and the measurement apparatuses and the only resources are the observed (nonlocal) correlations. Similarly, in 1SDI scenarios, where only one party (Bob) is trusted, it is natural to identify the resources for information processing tasks with the ensemble of states Bob obtains as a consequence of Alice's measurement (see also Ref.~\cite{Gallego14} for a discussion of this point).

Taking the above perspective, we are able to prove that any steerability problem can be translated into a joint measurability problem,  and vice versa. 
This result connects the well-known theory of joint measurements \cite{Busch85,Busch86} and uncertainty relations \cite{BLW14,UPrev,BLW13,BLWRev} to the relatively new research direction of steering. This is done by mapping any state ensemble for Bob in a corresponding steering-equivalent positive operator valued measure (POVM).  This simple technique is shown to give an intuitive way of generalizing the known results \cite{Uola14,Quint14}. Moreover, the power of the technique is demonstrated by mapping joint measurement uncertainty relations \cite{BLW14} into steering inequalities, and discussing the role of known steering monotones as monotones for incompatibility. 

{\it Preliminary notions.---} 
Given a quantum state $\rho$, i.e., a positive operator with trace one, an ensemble $\mathcal{E}=\{\rho_a\}$ 
for $\rho$ is a collection of positive operators such that $\sum_a \rho_a=\rho$. An assemblage $\mathcal{A}=\{\mathcal{E}_x\}_x$ is a collection of ensembles for the same state $\rho$, i.e., $\sum_a \rho_{a|x} =\rho$, for all $x$.
Similarly, a measurement assemblage $\mathcal{M}=\{M_{a|x}\}_{a,x}$ is a collection of operators $M_{a|x}\geq 0$ such that $\sum_a M_{a|x}=\openone$ for all $x$. Each subset  $\{M_{a|x}\}_a$  is called a positive-operator-valued measure (POVM), and it gives the outcome probabilities for a general quantum measurement via the formula $P(a|x)=\tr[M_{a|x} \rho]$.

A measurement assemblage $\mathcal{M}=\{M_{a|x}\}_{a,x}$ is defined to be {\it jointly measurable} (JM) \cite{Alietal09} if there exist numbers $p_M(a|x,\lambda)$ and positive operators $\{ G_\lambda\}$ such that 
\begin{equation}\label{eq:JM}
M_{a|x} = \sum_\lambda p_M(a|x,\lambda)\ G_\lambda,
\end{equation}
with $\sum_\lambda G_\lambda =\openone$, $p_M(a|x,\lambda)\geq 0$, and $\sum_a p_M(a|x,\lambda) =1$. Physically, this means that all the measurements in the 
assemblage can be measured jointly by performing the measurement $\{G_\lambda\}$
and doing some post-processing of the obtained probabilities.

In a steering scenario, a bipartite state $\rho_{AB}$ is shared by Alice and Bob.
Alice performs measurements on her system with possible settings $x$ and possible outcomes $a$, that is, the measurement assemblage $\{ A_{a|x}\}_{a,x}$. As a result of her measurement with the setting $x$, Bob obtains the reduced state $\varrho(a|x)$ with probability $P(a|x)$. Such a collection of reduced states and probabilities defines the state assemblage $\{\rho_{a|x}\}_{a,x}$, where
\begin{equation}
\rho_{a|x} = \tr_A[(A_{a|x}\otimes \openone) \rho_{AB}],
\end{equation}
with $P(a|x)= \tr [(A_{a|x}\otimes \openone) \rho_{AB}]=\tr_B[\rho_{a|x}]$ and $\varrho(a|x) = \rho_{a|x}/ P(a|x)$.
In particular, elements of the assemblage satisfy
\begin{equation}\label{redu}
\rho_B = \sum_a \rho_{a|x} = \sum_{a'} \rho_{a'|x'}, \text{ for all settings } x,x',
\end{equation}
where $\rho_B=\tr_A [\rho_{AB}]$. This expresses the fact that
Alice cannot signal to Bob by choosing her measurement $x$.

A state assemblage $\{\rho_{a|x}\}_{a,x}$ is called unsteerable if there exists a local hidden state (LHS) model, namely, numbers 
$p_\rho(a|x,\lambda)\geq 0$ and positive operators $\{ \sigma_\lambda\}$  such that
\begin{equation}\label{eq:LHS}
\rho_{a|x} = \sum_\lambda p_\rho(a|x,\lambda)\ \sigma_\lambda,
\end{equation}
with $\tr[\sum_\lambda \sigma_\lambda]=1$. A state assemblage is called
steerable if it is not unsteerable. The physical interpretation is the 
following: If the assemblage has a LHS model, then Bob can interpret
his conditional states $\rho_{a|x}$ as coming from the pre-existing states
$\sigma_\lambda$, where only the probabilities are changed due to the knowledge
of Alice's measurement and result. Contrary, if no LHS model is possible, 
then Bob must believe that Alice can remotely {\it steer} the states in his 
lab by making measurements on her side.

{\it Steerability as a joint-measurability problem.---} 
We now prove the main results of the paper, namely, that the steerability properties of a state assemblage can always be translated in terms of joint measurability properties of a measurement assemblage.

Let $\{\rho_{a|x}\}_{a,x}$ be a state assemblage and $\rho_B$ the corresponding total reduced state for Bob. We define ${\Pi_B: \mathcal{H}_B \rightarrow \mathcal{K}_{\rho_B}\subset \mathcal{H}_B}$ as the projection on the subspace $\mathcal{K}_{\rho_B} := {\rm range}(\rho_B)$, i.e., $\Pi_B \Pi_B^*=\openone_{\mathcal{K}_{\rho_B}}$ and $\Pi_B^* \Pi_B$ is a Hermitian projector in $\mathcal{L}(\mathcal{H}_B)$.  

Since $\rho_{a|x}$ are positive operators, Eq.~\eqref{redu} implies ${\rm range}(\rho_{a|x})\subset {\rm range}(\rho_B)$ for all $a,x$ \footnote{It is sufficient to notice that ${\rm Ker}(\rho_B)=\cap_{a,x}{\rm Ker}(\rho_{a|x})$ and ${\rm range}{A}={\rm Ker} A^\perp$ for any Hermitian operator $A$.}. Hence, we can define the restriction of our assemblage elements to the subspace $\mathcal{K}_{\rho_B}$ as $\tilde{\rho}_{a|x} = \Pi_B \rho_{a|x} \Pi_B^*$ and $\tilde{\rho}_B= \Pi_B \rho_B \Pi_B^*$, preserving the positivity of the operators. Such a restriction is needed in order to define $(\tilde{\rho}_B)^{-\frac{1}{2}}$ (see below).
Then, we define Bob's steering-equivalent (SE) observables $B_{a|x}\in \mathcal{L}(\mathcal{K}_{\rho_B})$ as
\begin{equation}\label{bobobs}
B_{a|x} = (\tilde{\rho}_B)^{-\frac{1}{2}}\ \tilde{\rho}_{a|x}\ (\tilde{\rho}_B)^{-\frac{1}{2}}.
\end{equation}
These operators are clearly positive and, by Eq.~\eqref{redu}, $\sum_a B_{a|x}=\openone_{\mathcal{K}_{\rho_B}}$, hence $\{B_{a|x}\}_a$ forms a POVM. 
We can formulate the first equivalence:

\begin{theorem}\label{th:bobs}
The state assemblage $\{\rho_{a|x}\}_{a,x}$ is unsteerable if and only if the measurement assemblage $\{B_{a|x}\}_{a,x}$ defined by Eq.~\eqref{bobobs} is jointly measurable. 
\end{theorem} 

{\it Proof.} First, notice that it is sufficient to discuss the existence of a LHS model for $\{ \tilde{\rho}_{a|x}\}_{a,x}$. From Eqs. \eqref{eq:LHS} and \eqref{eq:JM}, one can easily see that from a LHS for $\{ \tilde{\rho}_{a|x}\}_{a,x}$ one can construct a joint observable for $\{B_{a|x}\}_{a,x}$ and viceversa. The corresponding LHS model and joint observable are obtainable via the relation
\begin{equation}
G_\lambda = (\tilde{\rho}_B)^{-\frac{1}{2}}\ \tilde{\sigma}_{\lambda}\ (\tilde{\rho}_B)^{-\frac{1}{2}},
\end{equation}
where $\tilde{\sigma}_{\lambda}$ denotes the elements of the LHS for $\tilde{\rho}_{a|x}$.  
$\hfill \Box$

The above theorem shows that every steerability problem can be recast as a joint measurability problem. The other direction is trivial, since every joint measurability problem corresponds, up to a multiplicative constant, to a steerability problem with $\rho_B=\openone/d$. We can then state the main result:

\begin{theorem}\label{th:equiv}
The steerability problem of any state assemblage $\{\rho_{a|x}\}_{a,x}$ can be translated into a joint measurability  problem for a measurement assemblage $\{ M_{a|x}\}_{a,x}$, and vice versa.
\end{theorem}
 
It is now interesting to discuss the interpretation of Bob's SE observables.
Let $\rho=\sum_{i,j=1}^n\lambda_i\lambda_j\ket{ii}\bra{jj}$ be a pure state on a finite-dimensional Hilbert $\mathcal{H}_A\otimes \mathcal{H}_B$, where $\{\ket{i}_A\}_{1}^{d_A}$,$\{\ket{i}_B\}_{1}^{d_B}$ are the local bases associated with the above Schmidt decomposition of $\rho$, $n\leq \min\{d_A,d_B\}$, $\lambda_l>0$, and $\tr[\rho]=\sum_i \lambda_i^2=1$.

The reduced states for Alice and Bob have in such basis an identical form, namely, $\rho_{X}=\sum_{i=1}^n \lambda_i^2 \ket{i}\bra{i}_X$ with $X=A,B$, hence their ranges, $\mathcal{K}_{\rho_A},\mathcal{K}_{\rho_B}$ are isomorphic through the obvious mapping $\ket{i}_A\leftrightarrow\ket{i}_B$. 
Using that, we can formally write
\begin{equation}
\label{schmidtn}
\begin{split}
&\rho_{a|x} =\tr_A[(A_{a|x}\otimes \openone)\rho]\\
&=\sum_{i,j=1}^n\lambda_i\lambda_j\bra{j} A_{a|x}\ket{i}
\ket{i}\bra{j}
=\rho_A^{1/2} A_{a|x}^t \rho_A^{1/2},
\end{split}
\end{equation}
recovering a similar relation as in Eq. \eqref{bobobs}. The only missing step is to invert the relation by projecting on $\mathcal{K}_{\rho_B}$ and writing the inverse $\rho_A^{-1/2}$. 
Hence, for any pure state, Theorem \ref{th:bobs} gives us a clear interpretation of Bob's SE observables that generalizes the result given in Refs.~\cite{Uola14,Quint14}, namely, that for Schmidt rank $d$ state it is sufficient for Alice to use non jointly measurable observables in order to demonstrate steering.

\begin{remark*}
For a pure bipartite state, in order for Alice to demonstrate steering, her observable must be not jointly measurable even when restricted to the subspace where her reduced state, $\rho_A$, does not vanish.
\end{remark*}
 
Notice that the above remark holds also for pure separable states, however, since the corresponding subspace $\mathcal{K}_{\rho_A}$ is one-dimensional, joint measurability of Alice's observables is always trivially achieved. 

For the case of mixed states, a straightforward generalization of the above argument, e.g., via convex combinations, is not possible. Hence, the physical interpretation of Bob's SE observable for mixed states remains an open problem.

{\it Steering inequalities.---} We use the above result to give new steering inequalities for an assemblage arising from two and three dichotomic measurements for Alice when Bob's system is a qubit. We begin with the assemblage arising from two dichotomic measurements.

Given the assemblage $\{\rho_{a|x}\}$, with $a=\pm $ and $x\in\{1,2\}$, written in terms of Pauli matrices $\vec\sigma=(\sigma_1,\sigma_2,\sigma_3)$ as
\begin{equation}
\rho_{\pm|x}= t^\pm_x \openone + \vec s^{\ \pm}_x \cdot \vec\sigma,
\end{equation}
with $\vec s_x^{\ \pm}=(s_{1x}^{\ \pm},s_{2x}^{\ \pm},s_{3x}^{\ \pm})$, the only nontrivial case corresponds to a reduced state $\rho_B=\sum_{a=\pm} \rho_{a|x}$ of rank 2, otherwise the total state would be separable. 

\begin{figure}[t!]
\includegraphics[width=.42\textwidth]{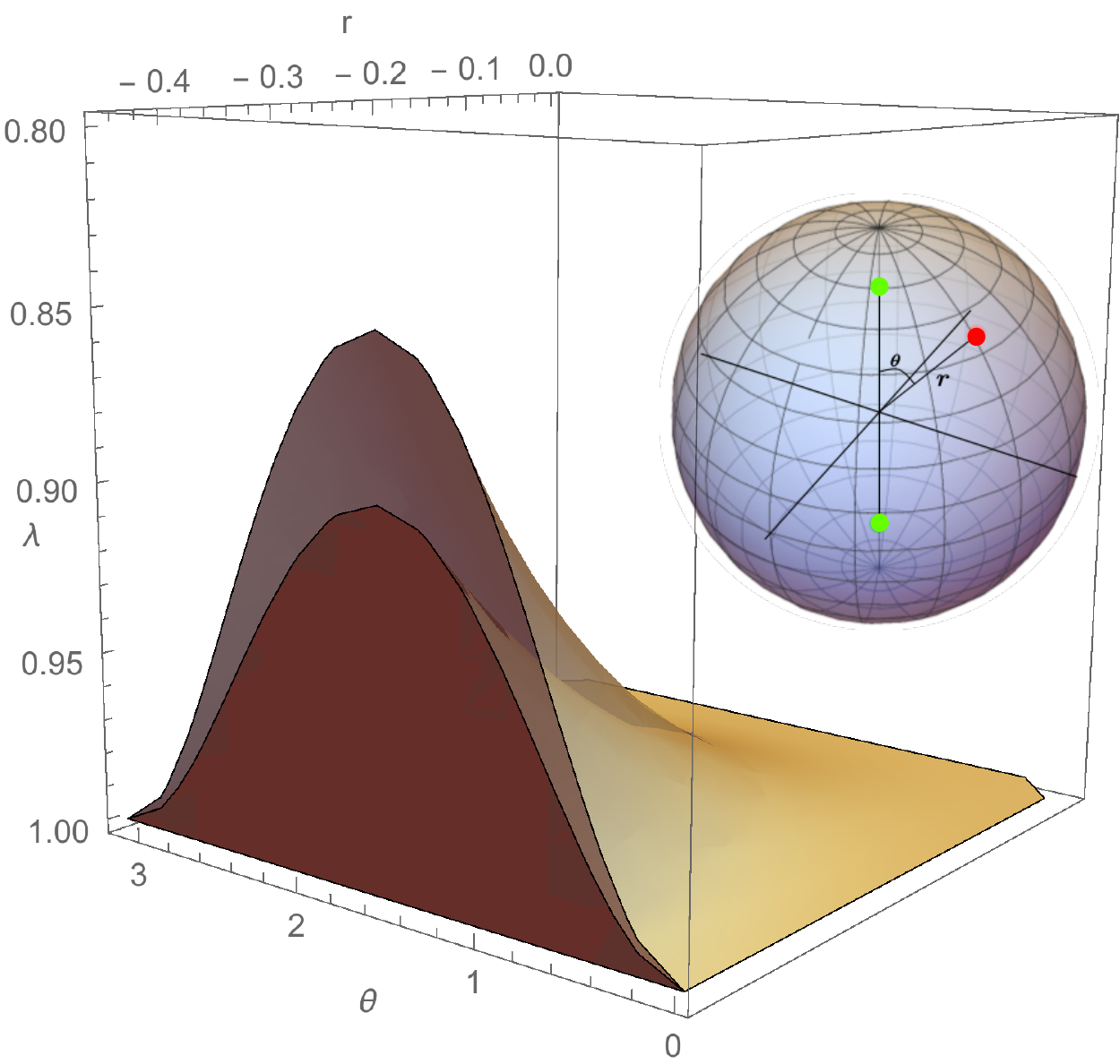}
\caption{\label{ineqplot}  Regions of the parameters $\lambda,r,\theta$ allowing for steering, detected by the inequality \eqref{SI2} (inner region) and inequality \eqref{Inequality1} (outer region), with $r=\|\vec{s}_2^{\ +}\|$ and $\theta$ the angle between $\vec{s}_2^{\ +}$ and the $z$ axis, and $t_2^+ = 0.45$ (fixed). { Inset}: representation in the Bloch sphere of the reduced states $\rho_{\pm|1}$ (green points) and $\rho_{+|2}$ (red point). The normalization factor $t_2^+=\tr[\rho_{+|2}]$ is not represented.}
\end{figure}

Then, the SE observables for Bob can be written as
\begin{eqnarray}\label{Bobobsqubit}
B_{+|x}=\frac{1}{2}((1+\alpha_x)\openone+\vec r_x\cdot\vec\sigma),\quad B_{-|x}=\openone - B_{+|x},
\end{eqnarray}
with $\alpha_x$ and $\vec r_x=(r_{1x},r_{2x},r_{3x})$ being functions of the assemblage $\{\rho_{a|x}\}$, the explicit forms of these functions is given in the the Supplemental Material. For such observables Busch {\it et al.} \cite{BLW14} have defined the {\it degree of incompatibility} to be the amount of violation of the following inequality
\begin{equation}\label{SI2}
\|\vec r_1 +\vec r_2\|+\|\vec r_1 -\vec r_2\|\leq 2.
\end{equation}
This inequality is a measurement uncertainty relation for joint measurements and as such it is a necessary condition for the joint measurability of two  observables on a qubit (see also Ref.~\cite{Busch86}). A violation of this inequality means that the SE observables of Bob are not jointly measurable and hence the setup is steerable. However, it has been shown that the degree of incompatibility does not capture all incompatible observables and a more fine-tuned version of this inequality, providing necessary and sufficient conditions has been derived \cite{YLO14}:
\begin{equation}\label{Inequality1}
(1-F_1^2-F_2^2)\left(1-\frac{\alpha_1^2}{F_1^2}-\frac{\alpha_2^2}{F_2^2}\right)\leq(\vec r_1\cdot\vec r_2-\alpha_1\alpha_2)^2, 
\end{equation}
with ${F_i=\frac{1}{2}(\sqrt{(1+\alpha_i)^2-\|\vec r_i\|^2}+\sqrt{(1-\alpha_i)^2-\|\vec r_i\|^2})}$, for $i=1,2$.

With the above definition, we can see the difference in the steerable assemblages detected by the steering inequality \eqref{SI2}, which provides only a necessary condition, and inequality  \eqref{Inequality1}, which completely characterizes steerability. Consider an ensemble of two reduced states along the $z$ axis and symmetric with respect to the origin, i.e., $\rho_{\pm|1}=\frac{1}{2}( \openone \pm \lambda \sigma_z)$. Given another ensemble $\rho_{\pm|2}$, by Eq.~\eqref{redu} only one of the two reduced states can be chosen freely, say $\rho_{+|2}=t_2^+ + \vec{s}_2^{\ +}\cdot\vec{\sigma}$, with the conditions $t_2^+\leq 1/2$ and $\|\vec{s}_2^{\ +}\|\leq t_2^+$. The steerability detected by Eqs.~(\ref{SI2}, \ref{Inequality1}) is plotted in Fig.~\ref{ineqplot}, for different values of the parameters $\lambda, r:=\|\vec{s}_2^{\ +}\|$, and the angle $\theta$ between $\vec{s}_2^{\ +}$ and the $z$ axis.

Finally, for the case of three dichotomic measurements on Alice's side (and Bob holding a qubit) we get three steering equivalent observables of the form Eq.~(\ref{Bobobsqubit}). For this case a joint measurement uncertainty relation and hence a steering inequality is given by \cite{YuOh13}
\begin{equation}\label{FTpoint}
\sum_{i=1}^4\|\vec R_i-\vec R_{FT}\|\leq 4,
\end{equation}
where $\vec R_1=\vec r_1+\vec r_2+\vec r_3$, $\vec R_i=2\vec r_{i-1}-\vec R_1$ $(i=2,3,4)$, and $\vec R_{FT}$ is the Fermat-Torricelli point of the vectors $\vec R_i$, i.e. the point which minimizes the left hand side of Eq.~(\ref{FTpoint}). Analogously to the case of Eq.~\eqref{SI2}, Eq.~\eqref{FTpoint} provides 
a necessary condition for the unsteerability of the state assemblage.

{\it Steering monotones.---} The previously known connection between joint measurability and steering \cite{Uola14,Quint14} has inspired the definition of incompatibility monotones, i.e., measures of incompatibility that are non increasing under local channels, based on steering monotones \cite{Pusey15} or associated with steering tasks \cite{Heinosaari15}.

Following the same spirit and in light of Theorem \ref{th:equiv},  we introduce a incompatibility monotone based on a recently proposed steering monotone, i.e., the {\it steering robustness} \cite{Piani15}. Given a measurement assemblage $\{ M_{a|x} \}_{a,x}$ we define the {\it incompatibility robustness} ($\mathcal{IR}$) as the minimum $t$ such that there exist another measurement assemblage $\{ N_{a|x} \}_{a,x}$ such that $\{ (M_{a|x} + t N_{a|x})/(1+t) \}_{a,x}$ is jointly measurable. The idea is to quantify the robustness of the incompatibility properties of the measurement assemblage under the most general form of noise. It is easily proven that $\mathcal{IR}$ can be computed as a semidefinite program and that it is monotone under the action of a quantum channel (cf. Supplemental Material).  

It is interesting to discuss the relation with previously proposed incompatibility monotones. In Ref.~\cite{Pusey15}, the {\it incompatibility weight} (IW), a monotone based on the {\it steerable weight} (SW) of Ref.~\cite{SNC14} was defined for a set of POVMs $\{M_{a|x}\}_x$ as the minimum positive number $\lambda$ such that the decomposition  $M_{a|x}=\lambda O_{a|x} + (1-\lambda) N_{a|x}$ holds for assemblage $\{N_{a|x}\}_{a,x}$ and jointly measurable assemblage $\{O_{a|x}\}_{a,x}$. From the definition it is clear that the IW suffer from a similar problem as SW, namely that whenever the elements of the (state or measurement) assemblage are rank-$1$, such weight is maximal. As a consequence, each pair of projective measurements, e.g., on a qubit, even along arbitrary close directions, are maximally incompatible according to IW, and, similarly, the state assemblage arising from a bipartite pure state, even with arbitrary small entanglement, is maximally steerable according to SW (see also the discussion in 
Ref.~\cite{Piani15}).

Another monotone has been proposed by Heinosaari {\it et al.} \cite{Heinosaari15}, based on noise robustness of the incompatibility with respect to mixing with white biased noise. This definition can be obtained from $\mathcal{IR}$, with the substitution $N_{a|x}\mapsto \frac{\openone}{d}$ (white noise) and,  for the corresponding coefficient $\lambda:= t/(1+t)$, the substitution $\lambda \mapsto (1+ab)\lambda$, in the case of dichotomic measurements, i.e., $a=\pm 1$. The notions of biasedness refers to the possibility of having a different disturbance for different outcomes.

\begin{figure}[t!]
\includegraphics[width=.42\textwidth]{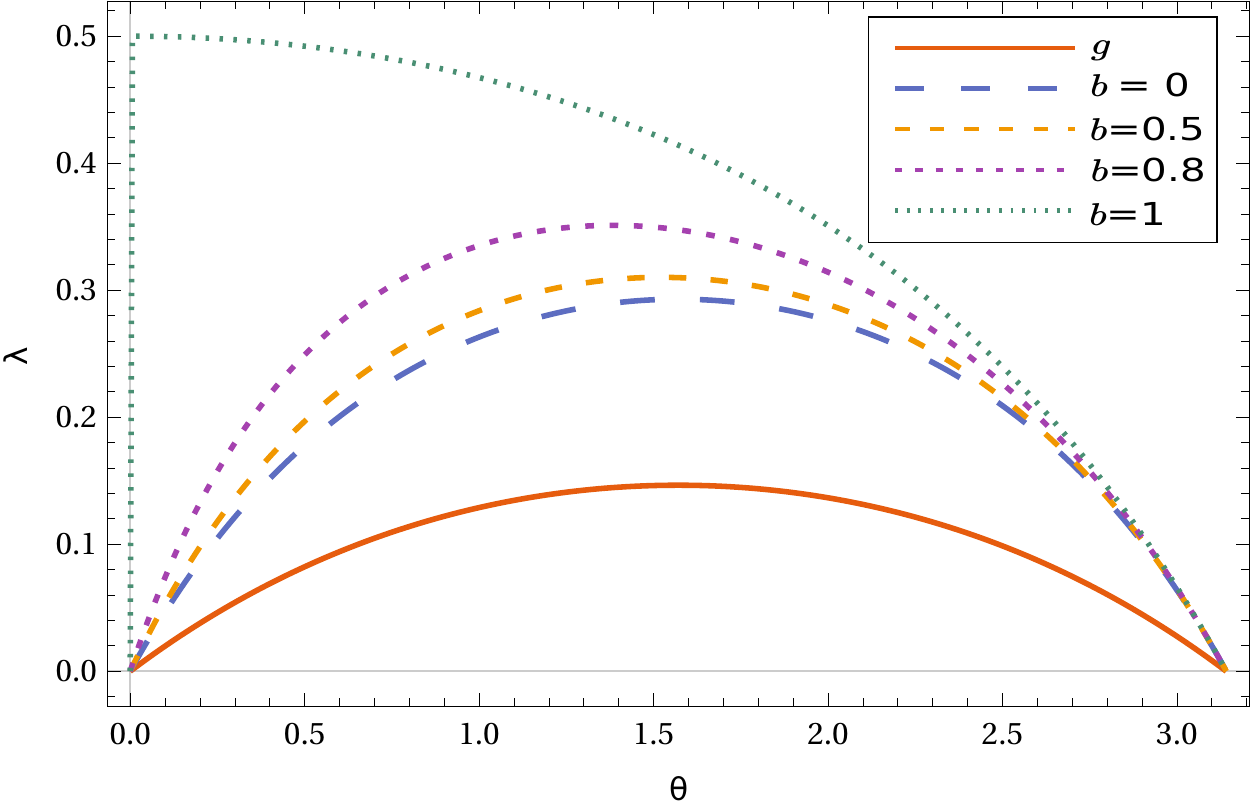}
\caption{\label{plotinc}Plot of noise robustness for white and general noise for two sharp qubit measurements separated by an angle $\theta$. The line denoted by $g$ corresponds to the parameter $\lambda_g$ of Eq. \eqref{mix1}, whereas lines denoted by $b$ to the parameter $\lambda_w$ of Eq. \eqref{mix2} for different level of bias, namely, $b=0, 0.5,0.8,1$ (see main text). The plot shows that the white noise tolerance is always at least double than the general noise tolerance $\lambda_g$. Moreover, the introduction of biased noise, quantified by the parameter $b$, with $b=0$ corresponding to unbiased white noise, only increases the noise tolerance.}
\end{figure}

As a consequence, $\mathcal{IR}$ is always a lower bound to the white noise tolerance. It is interesting to discuss such differences in a simple example. Consider a mixing of a measurement assemblage $\{M_{a|x}\}_{a,x}$ with white or general noise
\begin{eqnarray}\label{mix1}
\mathcal{M}_g&=&\{(1-\lambda_g) M_{a|x}+ \lambda_g N_{a|x} \}_{a,x},\  \\
\label{mix2} \mathcal{M}_w&=&\{(1-\lambda_w)M_{a|x}+\lambda_w \frac{\openone}{d} \}_{a,x}  .
\end{eqnarray}
If we choose in a qubit case $M_{a|x}=\frac{1}{2}(\openone+\vec v_{a|x}\cdot\vec\sigma)$ and $N_{a|x} =\frac{1}{2}(\openone-\vec v_{a|x}\cdot\vec\sigma)$ we end up with the mixings $\mathcal{M}_g=\{\frac{1}{2}(\openone+(1-2\lambda)\vec v_{a|x}\cdot\vec\sigma) \}_{a,x}$ and $\mathcal{M}_w=\{\frac{1}{2}(\openone+(1-\lambda)\vec v_{a|x}\cdot\vec\sigma) \}_{a,x}$.
It is then clear that in this case the noise robustness for general noise is always smaller than half the noise robustness with respect to white noise, namely,
\begin{equation}
\min \{\lambda_g | \mathcal{M}_g \text{  is JM }\}\leq \frac{1}{2}\min \{\lambda_w | \mathcal{M}_w \text{  is JM }\}.
\end{equation}
Explicit calculations (plotted in Fig.~\ref{plotinc}) show that the above choice for $N_{a|x}$ is not always the optimal one. The same noise robustness, for the case of orthogonal sharp measurements in dimension $d$, has been calculated in Ref.~\cite{Erkka}

The case of biased white noise corresponds to the substitution in Eq. \ref{mix2} $\lambda \mapsto \lambda(1+ab)$ for the case of binary measurements, i.e., $a=\pm 1$.
For the simplest case, i.e., two sharp projective measurement on a qubit, the noise robustness for  for mixing with general noise or with white noise plus a bias is plotted in Fig.~\ref{plotinc}.

{\it Conclusions.---} 
We have proven that every steerability problem can be recast as a joint measurability problem, and vice versa. As opposed to previous results \cite{Uola14, Quint14}, our approach does not include any assumption on the state of the system, but it is applicable knowing solely  Bob's state assemblage. This is arguably 
the most natural resource for steering, especially for  
one-sided device-independent quantum information protocols, 
where only Bob's side is characterized \cite{Gallego14}.

Our work connects the relatively new field of quantum steering 
with the much older topic of joint measurability. As we showed 
with concrete examples, that this connection allows to translate 
results from one field to the other. On the one hand, we were able 
to derive new steering inequalities for the two simplest steering 
scenarios based on joint measurability criteria for qubit observables. 
As opposed to previously defined steering inequalities based on SDP 
formulation \cite{Piani15,SNC14}, our inequalities are not defined 
in terms of an optimization for a specific assemblage, but are 
valid in general. For example, Eq.~\eqref{Inequality1} gives a 
complete analytical characterization of the simplest steering 
scenario for any state assemblage.

On the other hand, our result allowed to introduce a new 
incompatibility monotone based on a steering monotone. 
This opens a connection to entanglement theory: Similar
quantities as the incompatibility monotone 
have been used to quantify entanglement \cite{vidal1999,brandao2005,cavalcanti2006}. 
So, for future work it would be very interesting to use ideas 
from entanglement theory to characterize the incompatibility of 
measurements.

We thank M.~Piani and B.~A.~Ross for highlighting a problem (i.e., the lack of the normalization condition) in the initial definition of the SDP in Eq.~\eqref{e:SDP}. We thank F.~E.~S. Steinhoff and T.~Heinosaari for discussions and M.~C.~Escher for his help with 
Fig.~1. This work has been supported by the Finnish Cultural Foundation, the
EU (Marie Curie CIG 293993/ENFOQI), the FQXi Fund (Silicon Valley Community Foundation), and the DFG.

\appendix 
\section{Appendix}

\subsection{Explicit form of Bob's SE observables for a qubit and tight steering inequality}
Given the assemblage $\{\rho_{a|x}\}$, with $a=\pm $ and $x\in\{1,2\}$, written in terms of Pauli matrices $\vec\sigma=(\sigma_1,\sigma_2,\sigma_3)$ as
\begin{equation}
\rho_{\pm|x}= t^\pm_x \openone + \vec s^{\ \pm}_x \cdot \vec\sigma,
\end{equation}
with $\vec s_x^{\ \pm}=(s_{1x}^{\ \pm},s_{2x}^{\ \pm},s_{3x}^{\ \pm})$, the only nontrivial case corresponds to a reduced state $\rho_B=\sum_{a=\pm} \rho_{a|x}$ of rank 2, otherwise the total state would be separable. 

Since $\rho_B$ is full rank, we can directly compute first the square root $(\rho_B)^{\frac{1}{2}}$ and then its inverse $(\rho_B)^{-\frac{1}{2}}$ as a function of $\vec s_x^{\ \pm}$, either via a tedious direct calculation or with the aid of symbolic mathematical computation program.

Then the SE observables for Bob can then be obtained from the equation
\begin{equation}\label{bobobssm}
B_{\pm|x} = ({\rho}_B)^{-\frac{1}{2}}\ {\rho}_{\pm|x}\ ({\rho}_B)^{-\frac{1}{2}}.
\end{equation}
as
\begin{equation}\label{bobqubit}
B_{+|x}=\frac{1}{2}((1+\alpha_x)\openone+\vec r_x\cdot\vec\sigma),\   B_{-|x}=\openone - B_{+|x},
\end{equation}
with $\vec r_x=(r_{1x},r_{2x},r_{3x})$ and the substitutions
\begin{eqnarray}
\alpha_x &=& -1 + (2t^+_x \beta_0^2 -4 s_{3x}^{+} \beta_0 \beta_3 + 2t^+ _x\beta_3^2)/\Gamma^2,\\
r_{1x} &=& (2 s_{1x}^+\beta_1^2-4 s_{2x}^+\beta_1 \beta_2-2 s_{1x}^+ \beta_2^2)/\Gamma^2,\\
r_{2x} &=& 2(s_{2x}^+ \beta_1^2 + 2 s_{1x}^+ \beta_1 \beta_2 - s_{2x}^+ \beta_2^2)/\Gamma^2,\\
r_{2x} &=& 2(s_{3x}^+ \beta_0^2 + 2 t_{x}^+ \beta_0 \beta_3 - s_{3x}^+ \beta_3^2)/\Gamma^2,\\
\Gamma &=& (\beta_0^2 -|\vec{\beta}|^2),\\ 
\vec \beta &=& \frac{\lambda}{8\beta_0} (\vec s_{1}^{\ +} + \vec s_{2}^{\ +}),\\
\beta_0 &=& \frac{1}{2} \sqrt{ 1 - \sqrt{1-\lambda^2}},\\ 
\lambda &=& | \vec s_x^{\ +} + \vec s_x^{\ -}|.
\end{eqnarray}
Notice that $\lambda$ can be computed both from $\vec s_1^{\ \pm}$ and $\vec s_2^{\ \pm}$, it corresponds to the norm of the Bloch vector associated with Bob's reduced state.

\subsection{Incompatibility robustness as a semidefinite program}
The following construction is almost identical to the one presented in Ref.~\cite{Piani15}, we discuss it here for completeness. By definition
\begin{equation}
\begin{split}
 \mathcal{IR} = \min \Big\lbrace t \geq 0\  \Big|\ \frac{M_{a|x} + t N_{a|x}}{1+t}:=O_{a|x} \text{ are JM },\\
 \{ N_{a|x} \}_{a,x} \text{ measurement assemblage }  \Big\rbrace.
\end{split}
\end{equation}
We can then write 
\begin{equation}\label{Npos}
N_{a|x} = \frac{(1+t) O_{a|x} - M_{a|x}}{t} \geq 0, 
\end{equation}
where $\geq$ denotes a positive semidefiniteness condition. Eq. \eqref{Npos} is satisfied whenever
\begin{equation}
(1+t) O_{a|x} - M_{a|x} \geq 0,
\end{equation}
which can be rewritten, using the joint measurability properties of $\{O_{a|x}\}_{a|x}$, i.e.,  $O_{a|x} = \sum_\lambda p_M(a|x,\lambda) G_\lambda$ for all $a,x$, as
\begin{equation}
(1+t)\sum_\lambda p_M(a|x,\lambda) G_\lambda\geq M_{a|x} \ \forall a,x.
\end{equation}
By incorporating the factor $1+t$ in the definition of $G_\lambda$, one can easily see that the value of $1+\mathcal{IR}$ can be obtained via the following SDP:
\begin{equation}\begin{split}\label{e:SDP}
\text{minimize: } &\frac{1}{d}\sum_\lambda \tr[G_\lambda] \\
\text{subject to: }&
\sum_\lambda p_M(a|x,\lambda) G_\lambda\geq M_{a|x} \ \forall a,x, \\
 & G_\lambda\geq 0 .\\
 & \sum_\lambda G_\lambda = \openone \frac{1}{d}\left(\sum_\lambda \tr[G_\lambda]\right), 
\end{split}
\end{equation}
where the last equation encode the fact that $G$, up to the correct normalization, must be an observable. In addition, the postprocessing can be chosen, without loss of generality, as the deterministic strategy 
$p_M(a|x,\lambda) = \delta_{a,\lambda_x}$, where $\lambda:=(\lambda_x)_x$ and $\lambda_x$ is the hidden variable associated with the setting $x$, taking as value the possible outcomes $a$.

It can be easily proven that the program is strictly feasible (e.g., take $G_\lambda=\openone$) and bounded from below, i.e., the optimal value is 
always larger or equal one.

\subsection{Monotonocity of the incompatibility robustness under local channels}
To prove monotonocity of $\mathcal{IR}$ under the action of a quantum channel $\Lambda$ it is sufficient to prove that
\begin{equation}
\begin{split}
&\left\lbrace \frac{M_{a|x} + t N_{a|x}}{1+t}\right\rbrace_{a,x} \text{ is JM } \\
&\Longrightarrow \left\lbrace\Lambda \left( \frac{M_{a|x} + t N_{a|x}}{1+t}\right)\right\rbrace_{a|x}\text{ is JM }.
\end{split}
\end{equation}
Let us denote again $O_{a|x} := (M_{a|x} + t N_{a|x})/(1+t)$, with $\{O_{a|x}\}_{a,x}$ admitting a joint measurement, i.e., ${O_{a|x}=\sum_\lambda p_M(a|x,\lambda)\ G_\lambda}$. It is sufficient to check that $\{ \Lambda(O_{a|x})\}_{a,x}$ again admits a joint measurement ${\Lambda(O_{a|x})=\sum_\lambda p_M(a|x,\lambda)\ \Lambda(G_\lambda)}$. That $\Lambda(G_\lambda)$ is a POVM follows directly the properties of the channel $\Lambda$, since 
\begin{equation}
\begin{split}
&\Lambda(G_\lambda)\geq 0,\\ 
&\sum_\lambda \Lambda(G_\lambda)=\Lambda\left(\sum_\lambda G_\lambda\right)=\Lambda(\openone)=\openone.
\end{split}
\end{equation} 
Notice that, since we are looking for the transformation of the observables, we use the channel in the Heisenberg picture, hence the fact that the map is trace preserving when acting on states (Schr\"odinger picture) corresponds to its adjoint (Heisenberg picture) being unital.

\bibliography{steering-jm}{}

\end{document}